\documentclass[12pt]{iopart}
\usepackage{iopams}
\usepackage{setstack}

\usepackage[english,british]{babel}
\usepackage{graphicx}
\usepackage{hyperref}
\usepackage{epstopdf}
\usepackage{color}
\usepackage[utf8]{inputenc}

\usepackage{cite}
\newcommand{\ie}{i.\,e.\ }

\newcommand{\abs}[1]{|#1|}

\begin{document}
\title[Self-ordering in broadband radiation]
{Synthesizing variable particle interaction potentials via spectrally shaped spatially coherent illumination}
\author{D. Holzmann, M. Sonnleitner and H. Ritsch}
 \address{Institute for Theoretical Physics, University of Innsbruck, Technikerstra\ss e 25, A-6020 Innsbruck, Austria}
\ead{Daniela.Holzmann@uibk.ac.at}
\begin{abstract}
Collective scattering of spatially coherent radiation by separated point emitters induces inter-particle forces. For particles close to nano-photonic structures as, for example, nano-fibers, hollow core fibers or photonic waveguides, this pair-interaction induced by monochromatic light is periodic and virtually of infinite range. Here we show that the shape and range of the optical interaction potential can be precisely controlled by spectral design of the incoming illumination. If each particle is only weakly coupled to the confined guided modes the forces acting within a particle ensemble can be decomposed to pairwise interactions. These forces can be tailored to almost arbitrary spatial dependence as they are related to Fourier transforms with coefficients controlled by the intensities and frequencies of the illuminating lasers. We demonstrate the versatility of the scheme by highlighting some examples of unconventional pair potentials. Implementing these interactions in a chain of trapped quantum particles could be the basis of a versatile quantum simulator with almost arbitrary all-to-all interaction control. 
\end{abstract}
\section{Introduction}\label{intro}
Advances in laser cooling and manipulation of atoms and nano-particles nowadays allow one to prepare very low temperature ensembles where inter-particle light forces play an important role in the motional interaction. In 3D geometries the forces on two particles induced by collective light scattering or dipole-dipole interaction are typically of rather short range decaying as $(1/$distance$)^3$~\cite{brennen2000entangling}. However, by help of optical structures one can significantly increase the magnitude and range of light-mediated forces~\cite{stuart1998enhanced}. As the most prominent example single-mode high-$Q$ optical cavities implement resonantly enhanced infinite-range couplings~\cite{Fischer2001Collective}, which generate virtual all-to-all interactions and gives the basis for spontaneous crystallization into a regular lattice structure at high illumination power~\cite{ritsch2013cold}.

Similarly, in an alternative approach to enhance optical interactions one can use optical structures which tightly confine the radial extension of the electromagnetic field, allowing propagation only in one dimension. In the simplest example optical fibers can guide light virtually unattenuated over a very long range and consequently mediate optical forces over very long distances~\cite{holzmann2014self,ebongue2017generating}. For monochromatic illumination this can lead to an even more intriguing ordering phenomenon where light confines the particle motion and the particles confine the light along the fiber~\cite{griesser2013light}. 

Let us briefly recall the basic idea behind this here. Atoms trapped near a fiber or a 1D-photonic structure interact with the light field coupled into the fiber from its ends. This radiation field then travels along the chain of particles and encounters one atom after the other. Each atom then reflects and transmits some fraction of the light such that effective inter-particle forces emerge. However, as an alternative second method one can shine light transversely directly onto the atoms. Each atom then scatters some of this transverse pump field into the fiber where the light can travel and interact with the light field injected by other particles. Depending on whether the two fields interfere constructively or destructively the force on a pair of particles will be attractive or repulsive. This of course depends on the particle distance in units of the wavelength of the light. 

In previous works~\cite{holzmann2014self,holzmann2016tailored,ebongue2017generating,holzmann2015collective} we already demonstrated that such a configuration gives new tools to design particle-particle interactions. Especially the use of a broadband spectrum for the transverse pump beam gives finite-range forces which are very different from the interactions obtained in a longitudinally pumped fiber. Interestingly, the extreme case of very broadband black body radiation leads to very short range attraction and longer range repulsion.

In this work we go one step further and explore the effects of a transverse pump using multiple broadband beams of different central frequencies and bandwidths. As we will show in section~\ref{model} such a setup allows one to design almost arbitrary multi-particle interaction forces. Although these forces depend on the positions of each particle, to lowest order in the particle field coupling they can be written as a sum of two-particle interactions which in turn can be written as a Fourier series with coefficients given by the intensities of the transverse pump beams.

The option to design particle-particle interactions opens a host of possibilities and some of these shall be explored here in a generic way. In section~\ref{examples_free_particles} we discuss how this gives rise to new types of self-organisation. In section~\ref{examples_ext_potential} we explore opportunities for particles trapped in external potentials where tunable, position dependent interactions might be useful for quantum computation and simulation.

%
\section{Model} \label{model}
\begin{figure}
\centering
\includegraphics[width=0.8\textwidth]{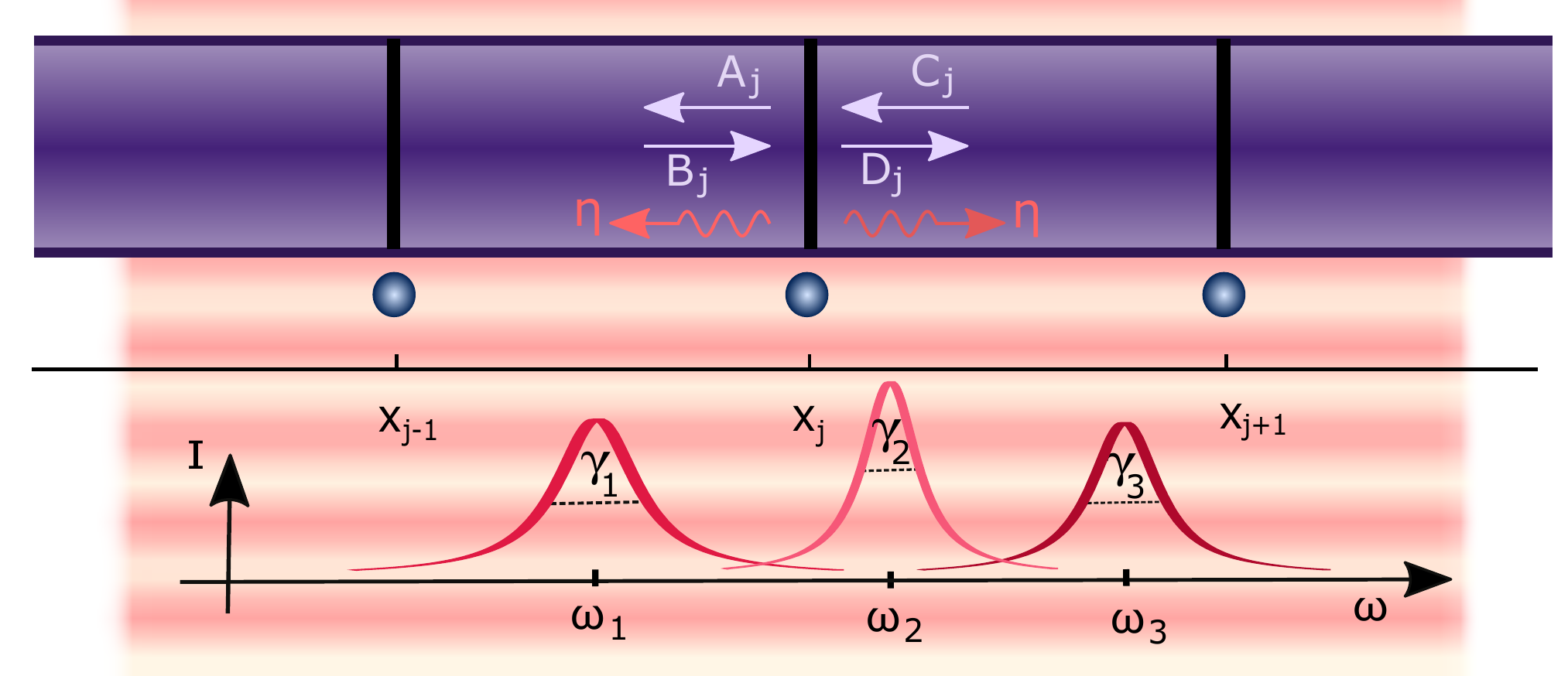}
\caption{A one-dimensional array of point particles scattering light into and out of an optical nano-structure can be modeled as a collection of beam splitters interacting with a plane wave. In this work we consider broadband transverse illumination characterized by the intensities, central frequencies and spectral widths of the incoming beams.}
\label{setup}
\end{figure}

Let us consider a set of linearly polarizable point-like emitters trapped in the evanescent field of light modes propagating in an effective 1D geometry. Typical examples are tapered nano-fibers, nano-waveguides or photonic crystal structures. As indicated in Fig.~\ref{setup} the particle motion is guided along the light propagation direction along which an additional lattice potential could be added, cf. section~\ref{examples_ext_potential}. When illuminated from the side the particles coherently scatter a fraction of the pump light into the guided modes characterized by an effective scattering amplitude $\eta$. Similarly some of the guided light is reflected into the reverse direction or scattered into free space. The injected photons propagate within the optical structure where they thus couple over very long distances to all other particles trapped near the fiber along their path. The particle-particle interaction mostly originates from the interference between the incoupled light amplitudes by the different particles. Closely related configurations have been experimentally realized with atoms~\cite{goban2012demonstration,vetsch2010optical,beguin2014generation,corzo2016large} or nano-particles~\cite{frawley2014selective}.

In the case of a single or a few particles a detailed numerical modeling of the setup is possible with good agreement to the experiment. However, as shown recently, the key physics can be already modeled and understood from a much simplified semi-analytical scattering model~\cite{maimaiti2016nonlinear}. We will generalize this model here towards multifrequency illumination and extract the key physics originating from frequency control.

As our toy model is, of course, strongly oversimplified, we will explicitly mention some of these simplifications here for clarity. We assume a scalar linear polarizability of the particles, while typical atoms have more than one or two contributing internal levels and thus exhibit a polarization, frequency and intensity dependent light shift. This is avoided using a $J=0$ to $J=1$ transition at very low magnetic field and light power. Similarly, the optical field around a nano-structure naturally exhibits radial and longitudinal polarization gradients, which sometimes generates even chiral couplings~\cite{petersen2014chiral}. This is ignored here but actually can be an extra asset to control the system, which we do not make use of here. We also simply assume a frequency independent polarizability and a flat fiber dispersion curve over a large frequency range, which is hardly the case. However, as only the product of polarizability times spectral density enters into the force, this poses no serious restriction to the generality of our model. As a further simplification we ignore radial forces acting against the transverse trap of the particles~\cite{maimaiti2016nonlinear}, which will give upper limits on the possible illumination powers, but generally do not seriously affect the longitudinal dynamics. 
    
\subsection{Beam-splitter model for transversely illuminated particles}

As we have outlined in a previous work~\cite{holzmann2014self} a set of particles next to a waveguide can be effectively modeled using a scattering matrix approach. Here for light propagating in the fiber, the particles simply act as beam splitters with a complex polarizability $\zeta $ with dispersive part $\Re(\zeta)$ and absorptive part $\Im(\zeta)$. The transmission and reflection coefficients $t=1/(1-i\zeta)$ and $r=i\zeta/(1-i\zeta)$ describe how the particles modulate the field inside the fiber. At the same time each illuminated particle also acts as a local source of light scattered into the propagating mode with (real) pump amplitude~$\eta$. The interaction between a particle and the fiber modes as well as the scattering into and out of the fiber can thus be effectively described by a $3\times 3$ beam splitter matrix $\mathbf{M}_{BS}$:
\begin{eqnarray}
\eqalign{\left( \begin{array}{c} A_j\\ B_j\\ \eta \end{array} \right)		
	&= \frac{1}{t}\left( \begin{array}{ccc} t^2-r^2 & r & \frac{1}{\sqrt{2}}( t-r) \\ -r & 1 & -\frac{1}{\sqrt{2}} \\ 0 & 0 & t  \end{array} \right)
				\left( \begin{array}{c}
				 C_j\\D_j\\ \eta \end{array} \right)
                 \\ 
				 &= \left( \begin{array}{ccc} 1+i\zeta & i\zeta & \frac{1}{\sqrt{2}}(1-i\zeta) \\ -i\zeta & 1-i\zeta & \frac{1}{\sqrt{2}}(i\zeta-1) \\ 0 & 0 & 1  \end{array} \right)
				\left( \begin{array}{c} C_j\\D_j\\ \eta \end{array} \right)
.}
\end{eqnarray}
Here $B_j,C_j$ are the amplitudes of the fields traveling within the fiber towards the $j$-th particle, while $A_j,D_j$ are the amplitudes of the outgoing fields. The parameter~$\eta$ gives the effective amplitude of the light scattered into the fiber and coming from the transverse pump field. For simplicity here we assume that all atoms see the same pump field amplitude and phase such that $\eta$ is the same for all particles. 

While the beam splitter relations above describe how the particles redistribute the light in and outside the fiber, a propagation matrix $\mathbf{M}_{P}(d)$ is needed to describe the propagation of light along the fiber between two particles with distance $d$,
\begin{equation}
\mathbf{M}_{P}(d)
	= \left( \begin{array}{ccc} e^{ikd} & 0 & 0\\ 0 &  e^{-ikd} & 0 \\ 0 & 0 & 1  \end{array} \right),
\end{equation}
with the wave vector $k=n(\omega)\omega/c$, frequency $\omega$, speed of light $c$ and refractive index $n(\omega)$, which we assume to be equal for all frequencies $n(\omega)\approx n$.

Multiplying the individual beam splitter and propagation matrices for a given particle distribution allows one to determine the field distribution along the fiber depending on the particle positions. The optical force on the $j$-th particle can then be calculated by inserting the calculated field amplitudes in the Maxwell stress tensor, which here simply gives~\cite{jackson1998classical,asboth2008optomechanical,deutsch1995photonic}
\begin{equation}\label{eq_force_BSj_general}
	F_j=\frac{\epsilon_0}{2}\left(\vert A_j\vert^2+\vert B_j\vert^2-\vert C_j\vert^2-\vert D_j\vert^2\right).
\end{equation}

The standard way to induce light forces on the particles is of course to send the light through the fiber. A standing light wave in the fiber creates thus a perfect one-dimensional lattice potential along the propagating directions. Note that the transverse field gradients can be also used to create trapping in the radial direction~\cite{corzo2016large,domokos2001efficient}. In the case of many particles trapped simultaneously, multiple scattering modifies this potential and already creates long range interactions and nonlinear dynamics~\cite{asboth2008optomechanical}. In general these interactions are tiny compared to the single particle potential as the backscattering per particle typically is small. However, when only a transverse pump is applied, all the light in the fiber is created from atomic scattering of the pump light. Here the atoms then interact not only via rescattering of this light, but even more prominently via interference of the light scattered by different atoms into the fiber~\cite{holzmann2014self}. In special cases this leads to ordering and self-organization of the atoms~\cite{griesser2013light} very similar as in transversely pumped cavities~\cite{ritsch2013cold}. But in contrast to a cavity there are no preselected modes defining the sensitivity of the system to the frequency of the incoming light.
 
\subsection{Forces due to transverse illumination}\label{}

The central aim of this work is to explore the possibilities to implement and design specific inter-particle forces induced by scattering of spatially coherent transverse pump fields with a variable spectrum. The forces stem from collective scattering into the fiber mode on the one hand and multiple rescattering of this light within the mode on the other hand. To simplify our model here we assume that the coupling of the particles to the fiber field is very weak, \ie $\abs{\zeta}\ll 1$, so that higher order scattering will be negligible. As shown in a previous work~\cite{holzmann2016tailored} one can even assume the limit $\zeta\rightarrow 0$ with only small error. Although the two coupling strengths $\eta$ and $\zeta$ both equally depend on the polarizability of the particles~\cite{masalov2013pumping} and the evanescent guided mode amplitude at their position, both parameters can be adjusted independently as only $\eta$ is proportional to the pump amplitude. Hence $\eta$ dominates for strong transverse pump compensating the weak coupling. 

In the limit $\zeta\rightarrow 0$ the forces on the particles can then be understood as a sum of long range two-particle interactions between all particle pairs at their positions $x_{l}$ and $x_j$ with $x_1 < x_2 < \dots <x_N$ and the force can be considered as a sum of contributions from different frequency components. As derived in some more detail in a previous work~\cite{holzmann2016tailored} for the spectral force density on the $j$-th of $N$ particles, i.e. the force contribution at frequency $\omega$ with intensity $I(\omega)$, we then get
\begin{eqnarray}
\hspace{-2.5cm}\eqalign{f_{j,N}(\omega,x_1,x_2,\dots,x_N)&=\frac{I(\omega)}{c}\left(\sum _{l=j+1}^{N} \cos \left(\frac{n\omega (x_{l}-x_j)}{c}\right)-\sum _{l=1}^{j-1} \cos \left(\frac{n\omega (x_{l}-x_j)}{c}\right)\right)\\
&=\sum _{l=j+1}^{N} f_{1,2}(\omega,x_{j},x_l)+\sum _{l=1}^{j-1} f_{2,2}(\omega,x_{j},x_l).}
\label{FomegaN}
\end{eqnarray}
Here $I(\omega)$ is the intensity distribution of the transverse pump field and the total force on this particle is then given by $F_{j,N}(x_1,\dots,x_N) := \int f_{j,N}(\omega,x_1,\dots,x_N) d\omega$.

The second line of Eq.~\eref{FomegaN} can be understood as follows: $f_{1,2}(\omega,x_j,x_l)$ gives the force density on the first of two particles located at $x_j$ and $x_l$ with $j<l$. Likewise $f_{2,2}(\omega,x_j,x_l)$ gives the force on the second particle of the pair, with $f_{2,2}(\omega,x_j,x_l) = - f_{1,2}(\omega,x_j,x_l)$. Hence the first sum describes the pairwise interaction with all particles to the right of $x_j$ while the second sum gives the interaction with all particles to the left. 

For the sake of analytical integrability we will assume the incident light field $I(\omega)$ as a sum of single transverse mode Lorentzian spectral lines with different central frequencies $\omega_m=2\pi c/ \lambda_m$, wave lengths $\lambda_m$, peak intensities $I_{m}=\vert\eta_m\vert^2 c\epsilon_0/2$ and widths $\gamma_m$:
\begin{equation}
I(\omega)=\sum_m\frac{I_{m}}{\pi}\frac{c\gamma_m}{c^2\gamma_m^2+n^2(\omega-\omega_m)^2}.
\label{i0}
\end{equation}
We also ignore here the frequency dependent polarizability of the scattering amplitude $\eta$ as this can be incorporated in a suitable frequency dependent effective pump strength. 

Integrating the force density Eq.~\eref{FomegaN} using this intensity distribution gives
\begin{eqnarray}
\eqalign{F_{1,2}(x_{j},x_l) &= \sum_m\frac{I_{m}}{\pi} \int_{-\infty}^{\infty }\frac{\gamma_m}{c^2\gamma_m ^2+n^2(\omega -\omega_m)^2}\cos \left(\frac{n\omega (x_{l}-x_j)}{c}\right) d\omega\\
&=\sum_m\frac{I_{m}}{nc}e^{-\gamma_m \vert x_{l}-x_j\vert}\cos\left(k_m (x_{l}-x_j)\right),}
\label{F2eq}
\end{eqnarray}
where from inversion symmetry we see $F_{1,2}(x_{j},x_l)=-F_{2,2}(x_{j},x_l)$. Using the fact that the force for a larger set of $N$ particles can be described using only pairwise interactions, we can define $F_{1,2}(x_{j},x_l)= -F_{2,2}(x_{j},x_l) \equiv F_\mathrm{pair}(x_{l}-x_j)\equiv F_\mathrm{pair}(d_{jl})$, with distance $d_{jl}=\vert x_l-x_j\vert$. Note that in this case $F_\mathrm{pair}(d_{jl})$ describes the force on the particle on the left.

Let us mention again here that we completely neglect possible interference effects between distant spectral components of the beams as we assume that they are only spatially but not time coherent and quickly average out in time. By help of the beam splitter method developed for single frequencies we thus can directly calculate the forces for each frequency component separately and then add them up to obtain the full force on the particles. Note that this implicitly also assumes low particle velocities so that the Doppler shifts on reflection do not couple different parts of the spectrum.

In a previous work on this model assuming a single illumination line with a finite bandwidth~\cite{holzmann2016tailored} we could identify stable stationary configurations of the particles where all forces vanish. As a central property for such stable particle configurations the outermost particles arranged to form Bragg mirrors and act like an optical resonator to trap high field intensities and the other particles between them. This behaviour vanishes for a large enough absorptive part $\Im(\zeta)$, where no stable solutions can be found due to outward radiation pressure in the chain. In addition the key effect of a finite laser bandwidth resulted in forces with exponential decay over distance proportional to the inverse bandwidth of the pump. The interaction range between the particles thus can be controlled via the spectral width of the incoming field. By changing $\gamma$ from small to large values, the interaction can be tailored from infinite range to nearest-neighbour coupling. This leads to the expectation that a more general spectral control of the input field should give much wider opportunities to tailor interaction potentials. 

\subsubsection{Two-particle forces and potentials}\label{two}

The radiation field inside the fiber is the result of collective scattering by all~$N$ particles. As a result, the force on each of these particles depends on the positions of all $N$ particles. But as discussed above, the total force on the $j$-th of $N$ particles can be rewritten as a sum of two-particle interactions depending only on the distance between these two particles.

\begin{figure}
\centering
\includegraphics[width=\textwidth]{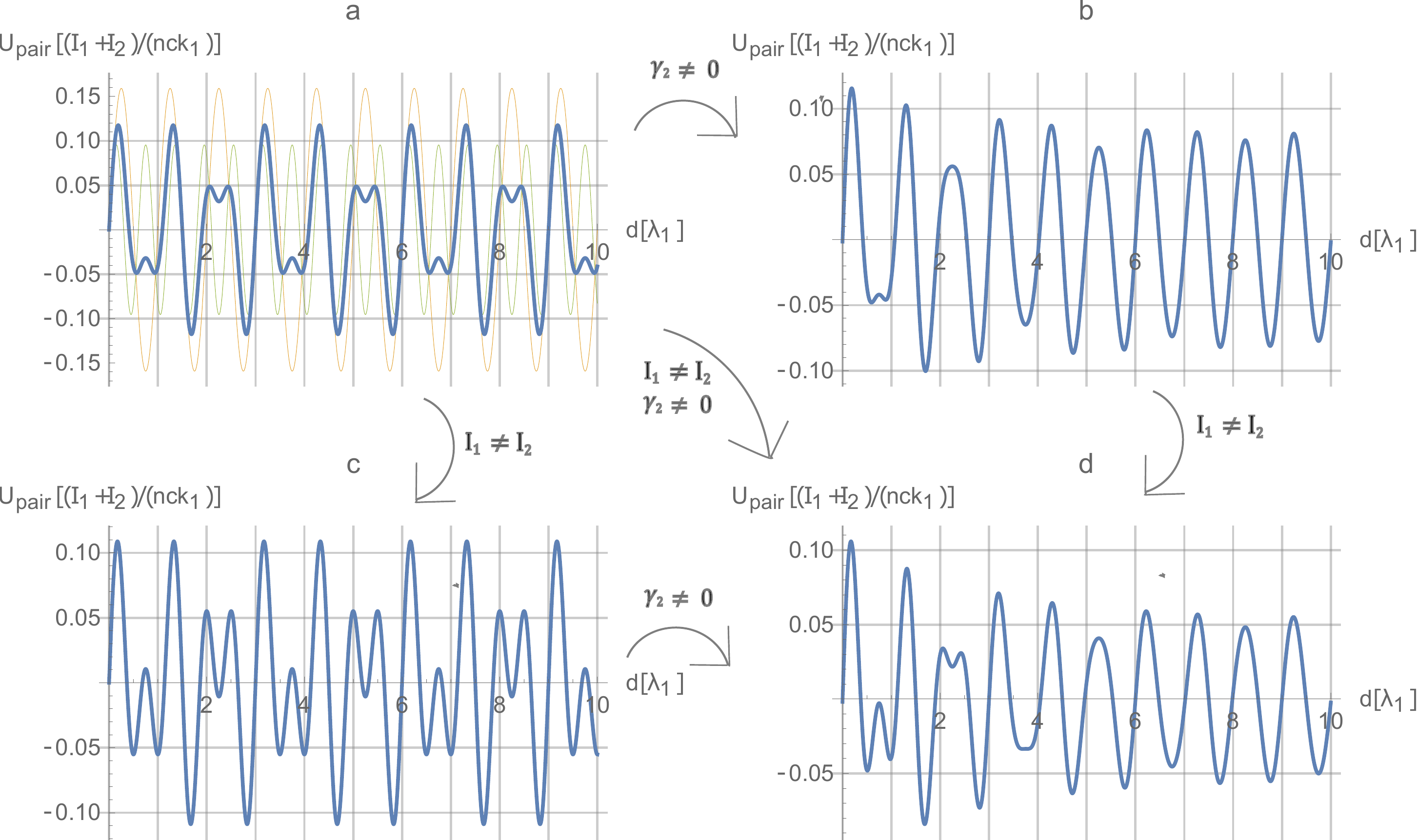}
\caption{Pair potential from Eq.~\eref{potential_2particles} for two particles illuminated by two fields with frequencies $k_1$ and $k_2=\frac{5}{3}~k_1$ as a function of the distance $d$. a) Adding two beams of equal intensity and without decay ($I_2=I_1$, $\gamma_1=\gamma_2=0$) creates a periodic potential landscape. Thin lines in the back show the individual fields with wave numbers $k_1$ (yellow) and $k_2=\frac{5}{3}~k_1$ (green) separately. Changing the intensity ratios and/or introducing finite bandwidths~$\gamma_{2}$ can create more complicate potentials: b) $I_2=I_1$, $\gamma_1=0$, $\gamma_2=0.05~k_1$, c) $I_2=2~I_1$, $\gamma_1=\gamma_2=0$, d) $I_2=2~I_1$, $\gamma_1=0$, $\gamma_2=0.05~k_1$.}
\label{F2part}
\end{figure}

The two-particle force given in Eq.~\eref{F2eq} can be expressed in terms of a potential, $F_\mathrm{pair}(d_{jl}) = -\partial_{x_j} U_\mathrm{pair}(d_{jl})$ with
\begin{equation}\label{potential_2particles}
	U_\mathrm{pair}(d_{jl})= \sum_m \frac{I_m e^{-\gamma_md_{jl}}}{nc(k_m^2+\gamma_m^2)}  \left( k_m \sin(k_md_{jl}) - \gamma_m \cos(k_md_{jl}) \right)
    \,.
    \label{poteq}
\end{equation}
Due to translational invariance of the setup, this potential depends only on the distance $d_{jl}=\vert x_l-x_j\vert$. The positions of global (local) minima of $U_\mathrm{pair}(d_{jl})$ gives the distances for stable (metastable) two-particle configurations. Clearly the potential is simply periodic for a single monochromatic transverse pump beam, but as shown in Fig.~\ref{F2part}, adding a second beam already gives intricate shapes for the pair-potential. Fig.~\ref{F2part} shows the potential for two particles illuminated by two broadband fields of varying intensity $I_2$ and width $\gamma_2$. We see that these two parameters as well as the frequency $\omega_2$ can be used to adjust the stable two-particle distances. The potential generated by a single beam has equidistant minima in the near field whose depth changes with $\gamma$ in the far field. For two fields we find non-equidistantly distributed equilibrium distances at short range. The relative intensity of the two fields can be used to adjust the relative depth of the potential wells. Note that in the chosen example no three minima of the potential have distances which are multiples of each other. So adding a third particle directly leads to frustration effects~\cite{kirkpatrick1977frustration,kim2010quantum} as the particles have to leave the stable pairwise configuration in order to find a configuration where all three particles are trapped. Finally, allowing for a finite range $\gamma$ of the second field as in Fig.~\ref{F2part}b) recovers the long range periodic potential with distortions resembling a defect at small distances.

\begin{figure}
\centering
\includegraphics[width=\textwidth]{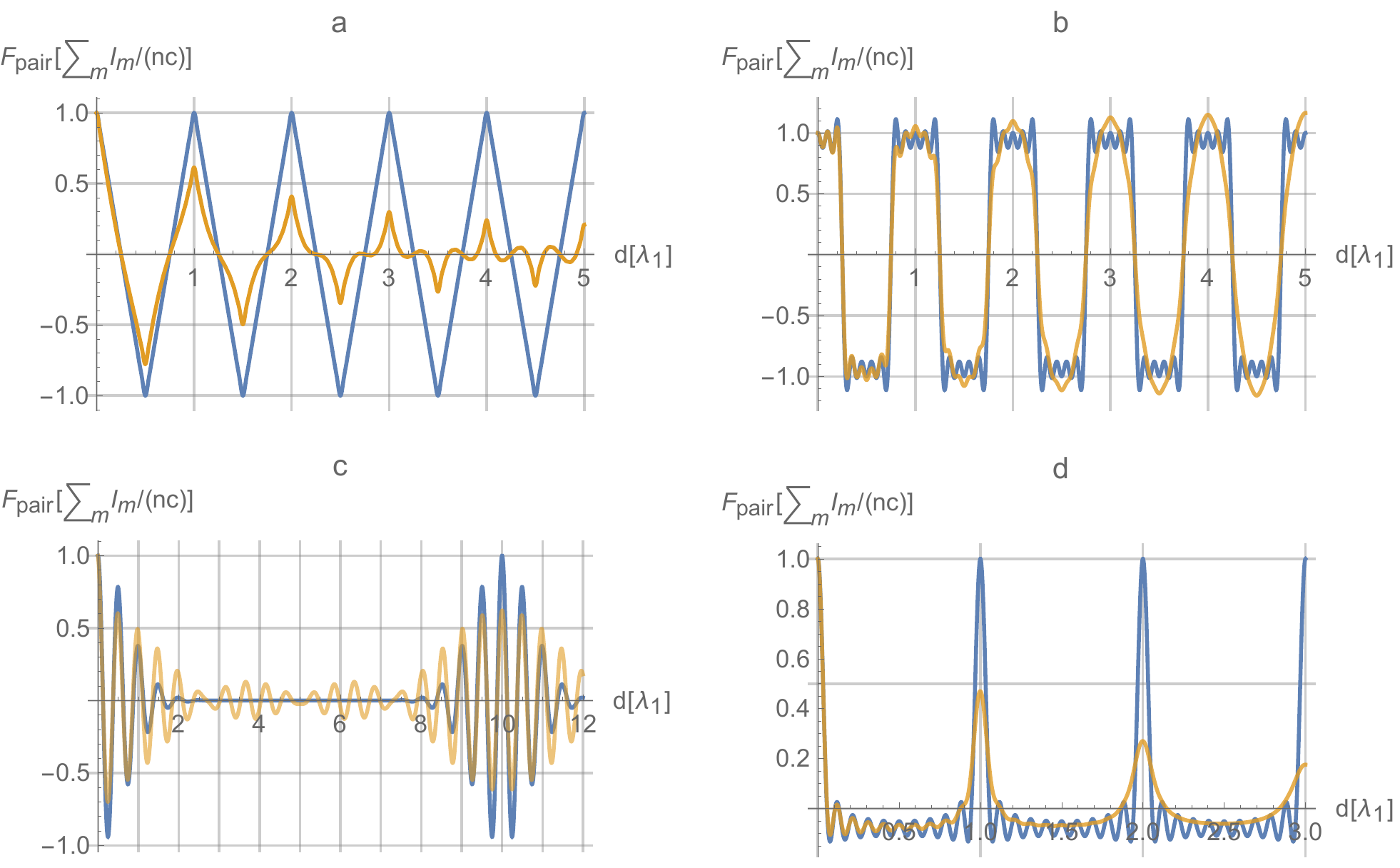}
\caption{Pair forces $F_\mathrm{pair}(d)$ of Eq.~\eref{F2eq} for two particles illuminated by a sum of $m_{\mathrm{max}}$ fields with wave numbers $k_1, \dots, k_{m_{\mathrm{max}}}$ as a function of the distance $d$ with $\gamma_m=0$ (blue) and $\gamma_m\neq 0$ (yellow). As the forces can be written as a Fourier series one can implement shapes including a) triangles, $I_m=\frac{1}{(2m - 1)^2}~I_1$, $k_m=(2m-1)~k_1$, $m_{\mathrm{max}}=10$, blue: $\gamma_m=0$, yellow: $\gamma_1=0.1~k_1$, $\gamma_{m>1}=0$; b) rectangular waves, $I_m=\frac{\sin(m \pi/2)}{m}~I_1$, $k_m=m~k_1$, $m_{\mathrm{max}}=10$, blue: $\gamma_m=0$, yellow: $\gamma_m=0.1(1-1/m)~k_1$; c) clusters of strong oscillating forces with a Gaussian intensity distribution $I_m=e^{-(m-10)^2/ 10}~e^{8.1}~I_1$, $k_m=(1+0.1m)~k_1$, $m_{\mathrm{max}}=20$, blue: $\gamma_m=0$, yellow: $\gamma_{m=\{1-8,13-20\}}=0.1~m~k_1$, $\gamma_{m=\{9-12\}}=0$; d) or very thin Lorentz peaks $I_m=(1-(m - 1)/10)~I_1$, $k_m=m~k_1$, $m_{\mathrm{max}}=10$ and blue: $\gamma_m=0$, yellow:  $\gamma_m=0.03~m~k_1$.}
\label{F2fourier}
\end{figure}

To further study the versatility of interaction potentials created by adding many frequency components we compare the expression for the two-particle force~\eref{F2eq} to the definition of a general Fourier series
\begin{equation}
f(t)=\frac{a_0}{2}+\sum_{m=1}^\infty \left(a_m\cos(mt)+b_m\sin(mt)\right),
\end{equation}
and see that the force~\eref{F2eq} is a special case of a Fourier cosine series with $a_0=b_m=0$, $a_m=\frac{I_{m}}{nc}$ and $\gamma_m=0$. Since $a_0=b_m=0$ we can only generate symmetric functions. Including a finite bandwidth $\gamma_m\neq 0$ this is no longer a classical Fourier series, but opens the possibility to change every single component of the series and design different force shapes as shown in Fig.~\ref{F2fourier}. When choosing specific combinations of $I_m$, $\omega_m$ and $\gamma_m$ we can model very peculiar force shapes such as, for example, a triangular wave. When changing $\gamma_m$ to non-zero values this triangle wave changes strongly in the far field (Fig.~\ref{F2fourier}a). Also a square wave can be modeled and when introducing $\gamma_m$ it changes to a simple sine wave (Fig.~\ref{F2fourier}b) in the far field. Fig.~\ref{F2fourier}c) shows an example with strong localizing forces appearing preferably around chosen distances. $\gamma_m$ can then be used to tune the forces between the peaks. An even more unusual case is a Lorentz distribution (Fig.~\ref{F2fourier}d) where very strong inter-particle forces appear specifically in a very narrow interval of distances. This includes a very sharp gradient of the force at the equilibrium positions providing for very strong particle confinement with deep potential wells.

Investigating the parameters more closely we can see that the examples given in Figs.~\ref{F2fourier}a), b) and d) rely on a setup spanning a very broad frequency range of several octaves: Starting with a far infrared beam (CO$_2$-laser) at ($\lambda_1=10\,000~nm$), the doubled frequency is at ($\lambda_2=5\,000~nm$) and even higher harmonics would reach the visible regime. Continuing this up to ten octaves more would thus certainly be hardly possible. The example of Fig.~\ref{F2fourier}c) shows a more accessible setup with a confined range of frequencies.

\subsubsection{Extension to many particles}

Summing up all pair-particle potentials Eq.~\eref{poteq} we can calculate the position dependent light shift (optical potential) experienced by a single particle keeping all the other particles at a fixed position as
\begin{equation}\label{potential_Nparticles}
	U_{j,N}(x_1,\dots,x_N) = \sum_{l=1}^{j-1} U_\mathrm{pair}(d_{jl})+\sum_{l=j+1}^{N} U_\mathrm{pair}(d_{jl})\,.
\end{equation}
Using $\partial_{x_j} U_\mathrm{pair}(d_{jl}) = -\partial_{x_l} U_\mathrm{pair}(d_{jl})$ we obtain the force acting on a particle of Eq.~\eref{eq_force_BSj_general} from $F_{j,N} = -\partial_{x_j} U_{j,N}(x_1,\dots,x_N)$. The total potential energy of the system can then be calculated by summing up all these potentials:
\begin{equation}\label{potential_Nparticles1}
	U_\mathrm{tot}(x_1,\dots,x_N)= \frac{1}{2}\sum_{j=1}^N U_{j,N}(x_1,\dots,x_N)
    \,.
\end{equation}

\section{Tailored forces for free particles and frustration effects of self-organization}\label{examples_free_particles}

As we have seen and illustrated in figure~\ref{F2fourier} there are many possibilities to tailor the forces acting between particle pairs. Using the fact that the equation for the force is a symmetric Fourier series we can produce every symmetric periodic function. Allowing for a bandwidth $\gamma_m$ for the various illumination lines allows to change the range of the individual frequency components tailoring the long range part of the force shape. In particular we can create forces with a very sharp gradient at the stable positions around zero force values, which implies narrow potential wells and very well localized configurations. We can also produce forces which only affect particles at certain chosen distances, while particles at other distances can move almost freely. 

\begin{figure}
\centering
\includegraphics[width=0.4\textwidth]{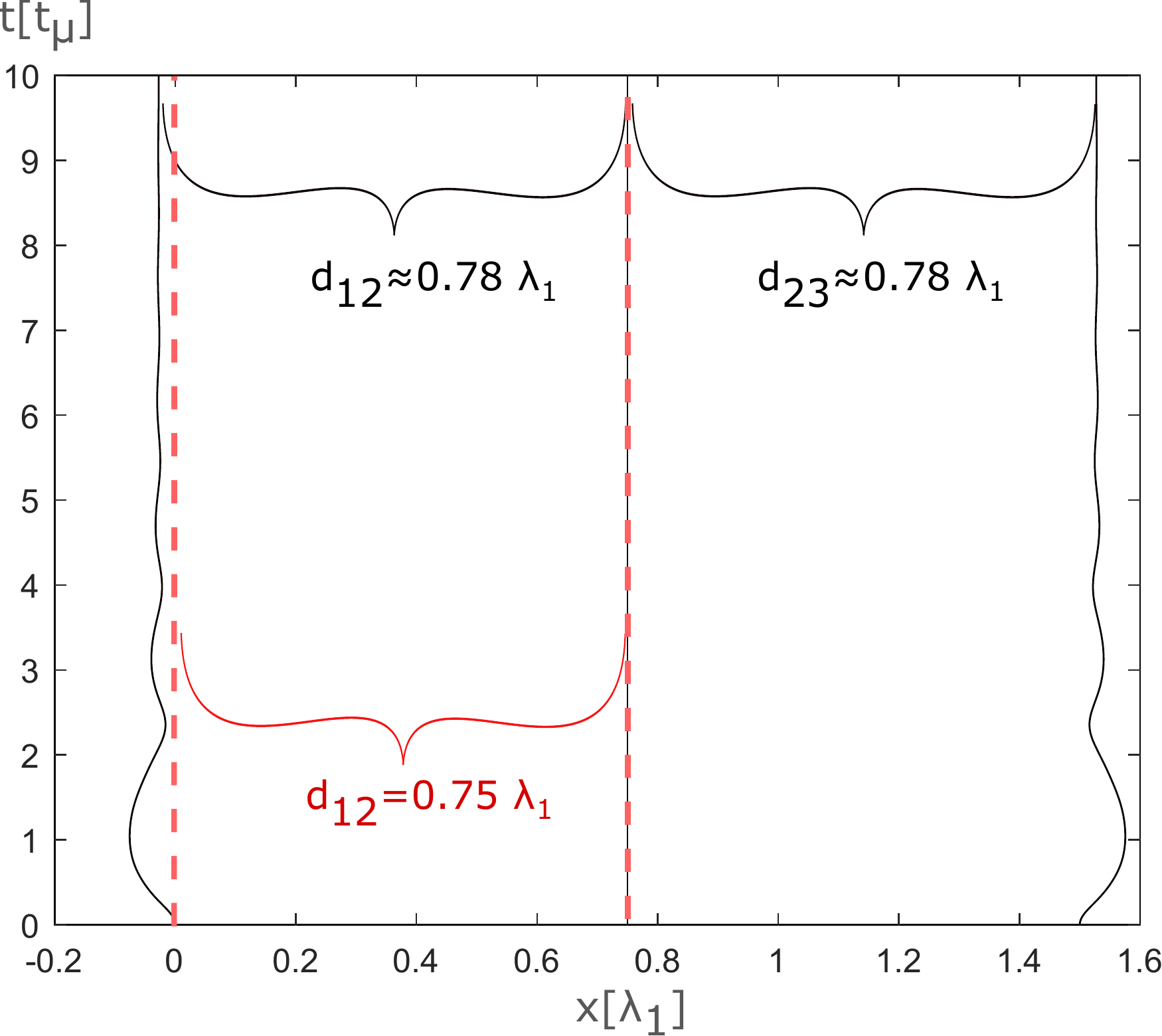}
\caption{Trajectories for three particles (black lines) illuminated by transverse pump-beams generating an approximately rectangular pair-force profile as shown in Fig.~\ref{F2fourier}b) with  $I_m=\sin(m \pi/2)/m~I_1$, $k_m=m~k_1$, $m_{\mathrm{max}}=10$ and $\gamma_m=0$. Red dashed lines show the case of two particles which remain at the minimum equilibrium distance. For this force pattern, three or more particles cannot arrange themselves such that the separation between each pair corresponds to a minimum of the respective pair-potential. The setup thus shows frustration as the particles have to reorder and find a shifted common equilibrium if a third particle is added. In this simulation we assume an environment with a friction coefficient $\mu$, the time time scale $t_\mu$ depends on $\lambda_1, I_m, n, \mu$ and the particle-mass $M$.}
\label{ffourier1}
\end{figure}

In figure~\ref{ffourier1} we show a set of three atoms arranging themselves under the influence of rectangular-wave two-particle forces as depicted in figure~\ref{F2fourier}b). Here, two particles order at a distance given by the very sharp zero points of the force, which is at $d_{12}=0.75~\lambda_1$ for the given parameters. If a third particle is added, each pair will seek to arrange itself at a distance given by the (stable) zero of the respective pair-force. But from fig.~\ref{F2fourier}b) we see that these stable distances are distributed such that this is impossible. The group of particles thus will have to ``compromise" and settle at distances $d_{12}=d_{23}\approx 0.78~\lambda_1$. This situation thus shows (classical) frustration~\cite{kirkpatrick1977frustration}, similar to magnets on a triangular grid which cannot arrange themselves such that each pair of magnets is in an anti-parallel configuration~\cite{kim2010quantum}.

%
\section{Designing interaction potentials between trapped particles}\label{examples_ext_potential}

\begin{figure}
\centering
\includegraphics[width=\textwidth]{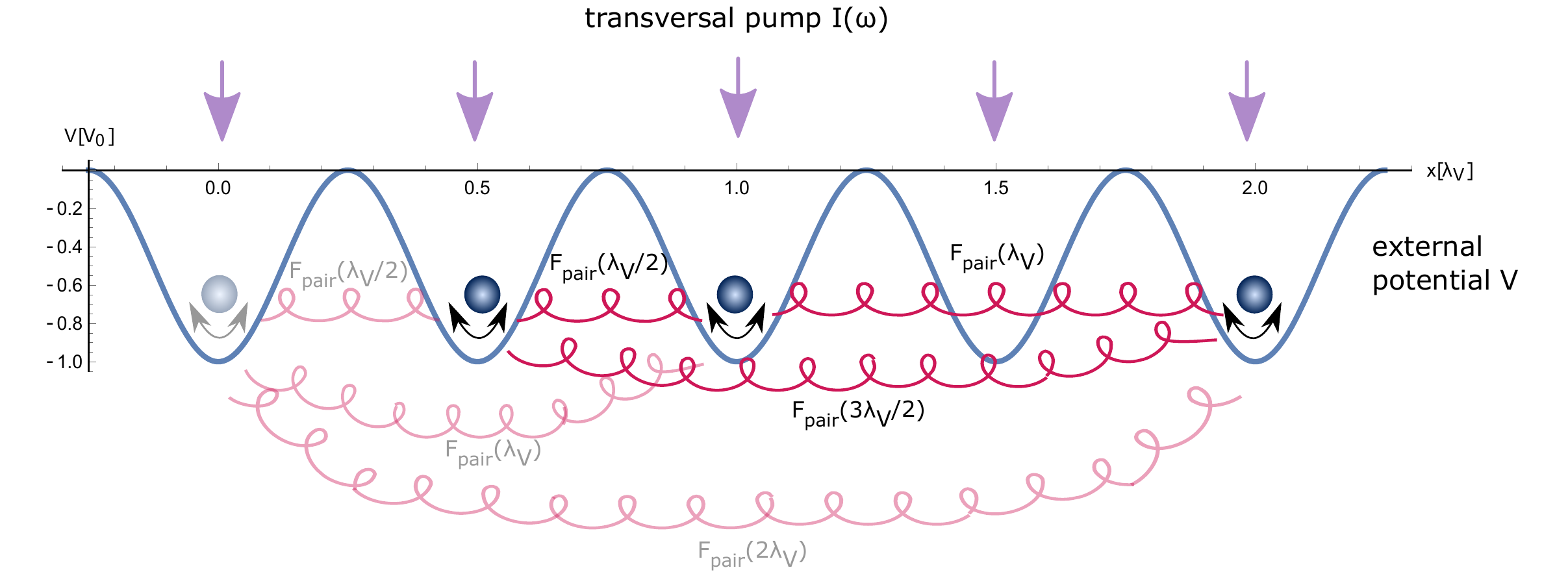}
\caption{Schematic illustration of particles trapped in an external periodic potential and illuminated by a transverse pump field. This transverse field leads to pair-interactions between the particles which slightly disturb the particles' trapping positions. Darker drawn lines correspond to the three particle setup used in Figs.~\ref{F34}a) and b). The configuration used in Fig.~\ref{F34}c) uses a fourth particle, indicated here with pale lines.}
\label{setuppot}
\end{figure}

In the previous section we discussed how free particles organize themselves and interact via tailored multi-particle forces generated by the scattered transverse light field. In many optical trapping applications, however, the particles are already trapped by a prescribed external lattice potential generated by standing wave light fields propagating through the nano-structure~\cite{corzo2016large,asenjo2017exponential}. As illustrated in figure~\ref{setuppot} the forces generated by the transverse pump field then provide a versatile tool to generate almost arbitrary interactions between particles trapped in the periodic minima of the external potential. Similar to the case of a multimode cavity this could be the basis of complex simulations~\cite{torggler2017quantum}.

We consider a strong periodic external potential
\begin{equation}\label{ext_potential}
V(x)=-V_0\cos^2(k_Vx),
\end{equation}
with $V_0=\frac{2I_V}{nk_Vc}$ depending on the intensity $I_V$ of the two incoming fields, $k_V$ the wave vector of the two fields, and examine the interaction between the particles due to the transverse pump field.

\begin{figure}
\centering
\includegraphics[width=0.9\textwidth]{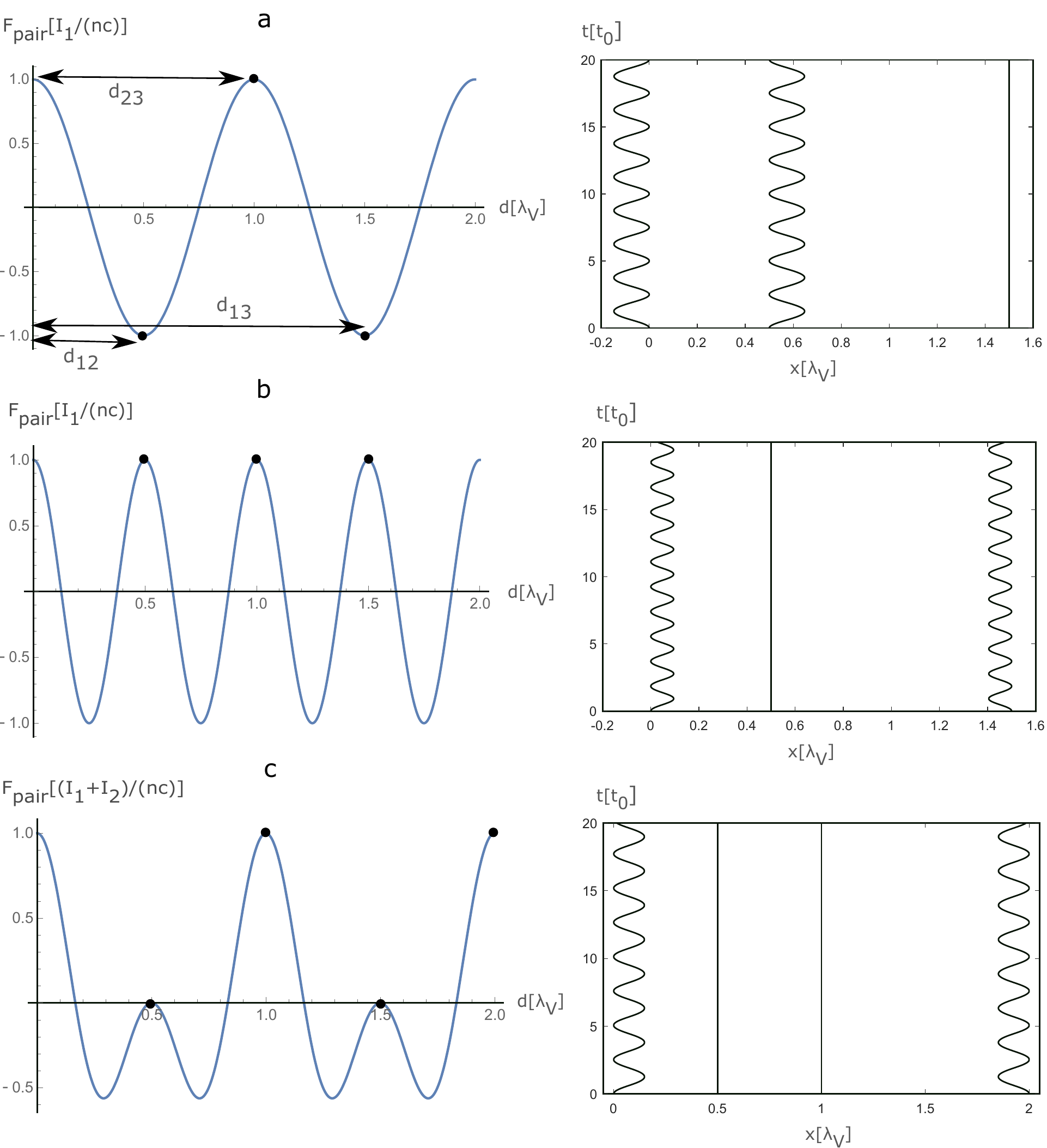}
\caption{The figures in the left column show distance-dependent forces between atom pairs for three different configurations of transverse fields and prescribed potentials, see also Fig.~\ref{setuppot}. The black dots indicate the respective pairwise separations of three (four) particles given by the periodic potential and chosen as starting points. In the figures on the right we show the corresponding trajectories of the particles reacting to these forces. We see that the pair-forces induced by the transverse beam induce coupled oscillation between a chosen pair of particles, while the forces conspire to leave the remaining particle(s) unaffected. The parameters here are $\gamma_1=\gamma_2=0$, $I_1=I_2$ for a) three particles, $k_1=k_V$, b) three particles, $k_1=2~k_V$, and c) four particles, $k_1=k_V,~k_2=2~k_V$.  The time-scale of the dynamics here is $t_0=\sqrt{Mnc(1/\sum_mI_m+1/I_V)}$. The two particle interactions shown in a) connect particle one and two, b) particle one and three, and c) particle one and four. It demonstrates how the single pair-forces can be combined to tailor the coupling between particles trapped in an additional periodic potential.}
\label{F34}
\end{figure}

For example, using an initial condition $d_{12}=0.5~\lambda_V$ and $d_{23}=\lambda_V$ places three particles in the first, second and fourth potential well, as depicted in Fig.~\ref{setuppot}. Adding transverse beams generates inter-particle forces which are given by the sum of pair-forces as shown in Eq.~\eref{FomegaN}. Fig.~\ref{F34} shows these pair-forces and the resulting trajectories of the particles. In Fig.~\ref{F34}a) we find that the total force on the third particle is the negative of the sum of the pair-force between the first and the third particle at a distance of $d_{13}=1.5~\lambda_V$ and between the second and the third particle at $d_{23}=\lambda_V$, which sum up to zero. So in this case only the first and the second particle interact. For Fig.~\ref{F34}b) we can observe the same for the first and the third particle. If we consider the four-particle case with the particles trapped in the first, second, third and fifth potential well as shown in~Fig.~\ref{setuppot}, the example~Fig.~\ref{F34}c) shows that by using two fields we can change the coupling between the particles and decouple the two particles in the middle.

\begin{figure}
\centering
\includegraphics[width=\textwidth]{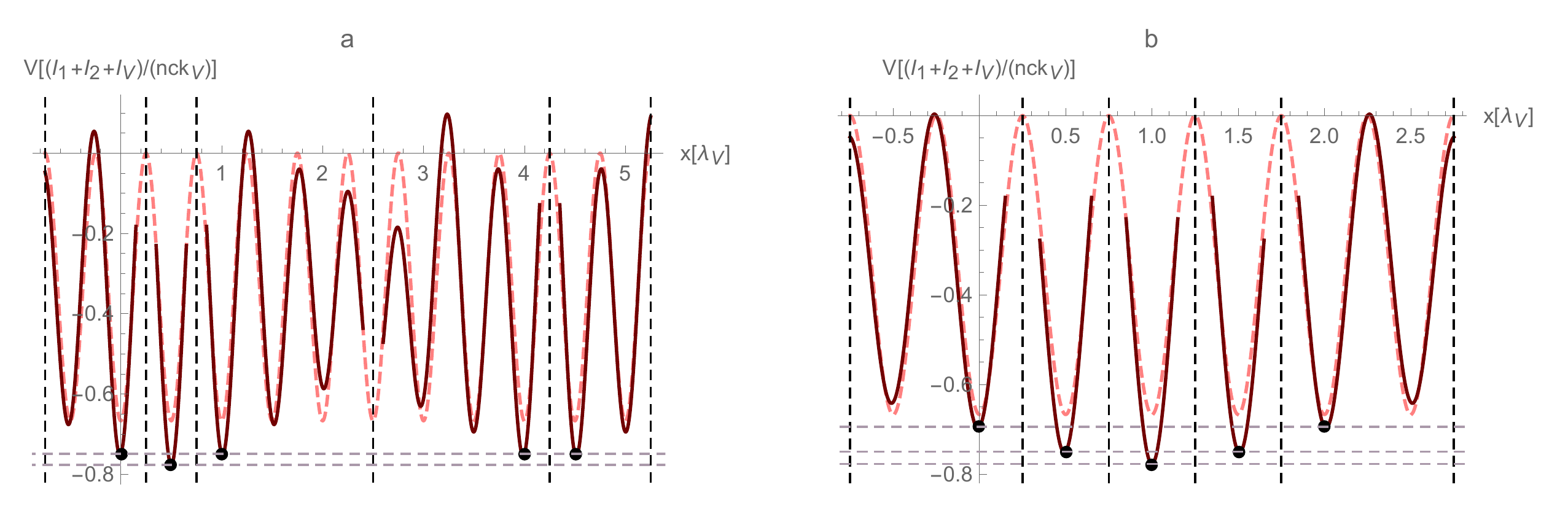}
\caption{\label{Vpot}%
Potential for five particles trapped in an external potential and interacting via two transverse fields with intensities $I_1=I_2=I_V$, wave numbers $k_1=k_V$, $k_2=5/3~k_V$ and spectral widths $\gamma_1=\gamma_2=0$. Red dashed lines show the external potential $V(x)$, black points the particle positions. For each particle the respective potential $U_{j,N}$ is drawn in solid red curves between the dashed vertical lines. That is, for example, between the first two vertical lines we show the potential $U_{1,N}$ if the position of the first particle is varied while the other particles are fixed at the positions marked by the black dots. The configuration of particles placed in the first, second, third, ninth and tenth well shown here is the one minimizing the total potential energy. A configuration where all particles are in neighbouring wells as shown on the right has a higher total energy.}
\end{figure}

In general the scattering forces on $N>2$ particles cannot be described by a common potential~\cite{asboth2008optomechanical} and the dynamics of the particles does not necessarily conserve energy as the pump provides for a non-depleting energy and momentum source. However, to lowest order in particle-field coupling as considered above, every particle pair has its own relative potential energy~$U_\mathrm{pair}(d_{jl})$ and in Eq.~\eref{potential_Nparticles} we gave the potential energy for each particle, if the positions of other particles are fixed. But changing the position of one particle affects the optical potential of all other particles. 

\begin{figure}
\centering
\includegraphics[width=\textwidth]{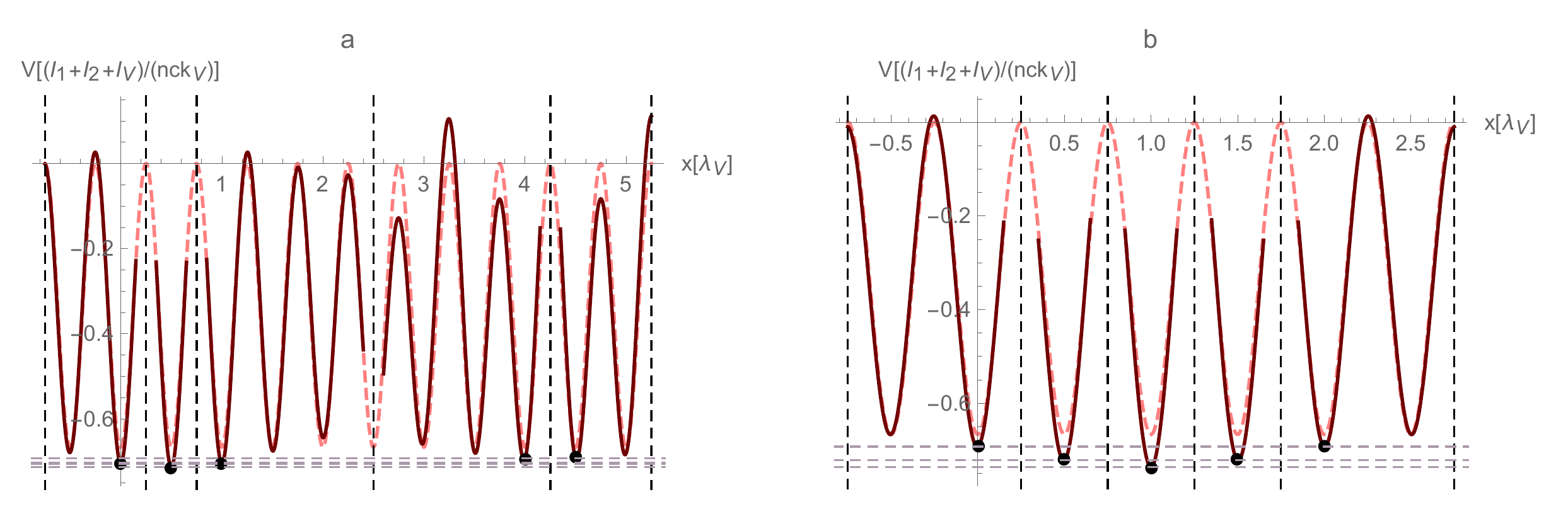}
\caption{\label{Vpotg}%
The same as in figure~\ref{Vpot}, but with a broad spectral width of the second transverse beam, $\gamma_2=0.1~k_V$. For these parameters we see that the particles minimize their collective total potential energy if they are in neighbouring wells as shown on the right hand side. Introducing $\gamma_2$ to the system thus modulates the long-range interactions between the particles, such that now the configuration on the left is less favourable.}
\end{figure}

But of course, for every set of parameters (the number of particles, strength of the external potential and settings for transverse beams) there will be a configuration of particle positions minimizing the total potential energy given in Eq.~\eref{potential_Nparticles1}. For $N$ particles, this is an $N-1$-dimensional optimization problem which is conceptually visualized in Figs.~\ref{Vpot} and~\ref{Vpotg}. There we show the sum of the external potential~$V(x)$ and single particle potentials $U_{j,N}$ for $N=5$ particles. For the chosen parameters Fig.~\ref{Vpot}a) shows that the lowest energy occurs where the particles are trapped in the first, second, third, ninth and tenth potential minima (indicated by the black dots). Comparing this to Fig.~\ref{Vpot}b) we find that the total potential is not minimized if the particles are placed in neighbouring wells.

The transverse beams in Fig.~\ref{Vpot} are chosen with a vanishing linewidth $\gamma_{1,2}=0$. But introducing a broad spectrum for the second laser, $\gamma_2=0.1~k_V$, changes the interactions in the far field such that placing the particles in neighbouring wells gives the lowest-energy configuration, as shown in Fig.~\ref{Vpotg}. 

Such a system thus offers the possibility to manipulate and rearrange particles trapped in an external potential. Changing the intensities, frequencies and widths of the fields we can suppress or push the individual two-particle-interactions. These controlled couplings then give the opportunity to realize a desired energy landscape useful, for example, in analog quantum simulation or quantum annealing~\cite{finnila1994quantum,georgescu2014quantum,torggler2017quantum}.

In a setup where the inter-particle forces are weaker than the strong external potential $V$ given in Eq.~\eref{ext_potential}, one can expand the position of the particles around the minima of the external trap, $x_j = x_j^0 + \Delta_j$ where $k_V x_j^0 = z_j \pi$ for an integer $z_j$ with $z_j>z_l$ if $j>l$. One can then write the total external potential as a sum of harmonic oscillators
\begin{equation}
\sum_{j=1}^N V(x_j) \simeq - \sum_{j=1}^{N} V_0\left(1-k_V^2 \Delta_j^2\right).
\end{equation}
Setting $\gamma_m=0$ we get for the pair interaction
\begin{eqnarray}
	\eqalign{U_\mathrm{pair}(d_{jl}) \simeq \sum_m \frac{I_m}{nck_m}&\Bigg(\sin(\vert z_j-z_l\vert \pi k_m/k_V) \left(1- \frac{k_m^2}{2} \left(\Delta_j-\Delta_l\right)^2\right)\\
&+ \cos(\vert z_j-z_l\vert \pi k_m/k_V) k_m \left(\Delta_j-\Delta_l\right)\Bigg).}
\end{eqnarray}
Tuning the frequency ratios $k_m/k_V$ thus allows one to design quadratic or linear interactions between each pair of particles while choosing a finite bandwidth $\gamma_m$ gives control over long-range coupling. One can imagine that this might be useful to implement specifically designed multi-particle Hamiltonians with coupled motional states of atoms trapped in specific distant lattice sites.

%
\section{Conclusions}
Cold atoms, molecules or point-like nano-particles trapped along light-guiding nano-structures constitute a powerful setup to study the physics of collective light scattering in general and long range light mediated inter-particle forces in particular. Several experiments have now entered the domain, where the contribution of even a single particle to the field scattered into and out of the confined modes is significant and directly observable~\cite{tiecke2014nanophotonic, goban2015superradiance}. So far experiments and virtually all theoretical approaches have concentrated on the action of monochromatic laser light coupled into single field modes. In this work, at the hand of a simplified effective model, we theoretically exhibit the wealth of new possibilities arising from applying a transverse pump field with a freely designable spectrum. Typically such a multispectral field can be composed by a set of independent spatially coherent laser beams of different central frequency and spectral width. Alternatively spectrally reshaped white light sources could be envisaged.  

We showed that in the limit of weak internal backscattering the two-particle forces arising in such a setup can be synthesized from different spectral components each contributing a component to a Fourier cosine series of the force with an amplitude proportional to the field intensity scattered into the fiber. Hence, in principle, one can almost arbitrarily shape the two-particle forces along one dimension and implement any real symmetric function of distance. As the force is based on coherent scattering it constitutes a conservative force without significant heating. Hence a generalization to trapped quantum particles for the implementation of a quantum simulator with virtually all to all coupling can be envisaged.    

At this point our simplified effective model exhibited very promising results which have to be confirmed in more realistic models. In particular, implementing the strong confinement and low temperatures of the transverse motion are very challenging although some experiments achieved close to ground-state cooling~\cite{goban2012demonstration}. Thus fluctuations of the coupling strength should be included. We also have ignored issues of polarization gradients and chirality, which add technical challenges, but also further enhance the versatility of the implementations~\cite{beguin2017observation, asenjo2017exponential}. 

In summary, although an actual physical implementation is challenging, the forces due to multimode transverse illumination provide a powerful and very versatile tool to explore many-particle self-organization, unconventional interactions and possibly also implement quantum simulations allowing to address individual particles in 1D optical lattices.

\ack
This work was funded by the Austrian Science Fund FWF grants No. F4013 SFB FOQUS and J3703.
%

%
\section*{References}

\end{document}